\documentstyle[aps,epsfig]{revtex}
\begin{document}
\begin{flushright}
hep-ph/9707529 \\
AS-ITP-97-16 \\
July 1997
\end{flushright}
\vspace*{1cm}

\begin{center}
{\Large  Polarized $\Lambda_b \rightarrow X_c \tau \nu$ in the SM and
THDM } \footnotetext{This work was supported in part by the National
Natural
Science Foundation of China and partly supported by Center of Chinese Advanced
Science and Technology(CCAST).}

\vspace*{1cm}
 Chao-Shang Huang$^{1,2}$ and Hua-Gang Yan$^{1}$

\vspace*{0.3cm}

$^{1}$ Institute of Theoretical Physics, Academia Sinica,\\
\hspace*{.3cm} Beijing 100080, China\\
$^{2}$ International Centre for Theoretical Physics,\\
\hspace*{.3cm} P.O.Box 586, 34100 Trieste, Italy
\end{center}
\parindent=1.2cm
\begin{center}
{\large \bf Abstract}\\
\vspace{.5cm}
\begin{minipage}{13cm}
{\small  The inclusive rate and  $\tau$ spectrum for a  polarized $\Lambda_b$
-baryon to decay to charm hadronic final states and leptons $\tau \nu$ in
the SM and a two-Higgs doublet model are computed.The $O(\alpha_s)$ QCD
corrections to $\tau$ spectrum in the two-Higgs model are also given.
}
\end{minipage}
\end{center}

\clearpage

\begin{center}  
{\large \bf I.Introduction}
\end{center}

\parindent=18pt
The semileptonic decay $B \rightarrow X_c l \nu$ has been extensively studied
in both the standard model(SM) and a two-Higgs-doublet
model(THDM)\cite{1,2,3,4,5,6,7,8,9}.Compared with the B decay,in addition to the spectrum of the
lepton arising from the decay,the various spin correlation quantities are of 
interest for the decay of a polarized $\Lambda_b$.It's well-known\cite{10,11} that 
the heavy quarks produced in $Z^0$ decay are polarized and only charmed and
beautiful $\Lambda$ baryons seem to offer a practical method to measure the 
polarization of the corresponding heavy quark.The polarization transferred
from a heavy quark Q to the corresponding $\Lambda_Q$ is $100\%$\cite{12} in the
limit $m_Q \rightarrow \infty$.Thus the angular distributions of charged 
leptons\cite{13,14} from semileptonic decays of $\Lambda_b$ and $\Lambda_c$ can be
used as spin analysers for the decays of heavy quarks.Some aspects of the inclusive rate
and 
$l$ spectrum of polarized $\Lambda_b \rightarrow X_c l \nu$ have been
studied in the SM\cite{3,k1,k2}.

The $\Lambda_b \rightarrow X_c \tau \nu$ (and $B \rightarrow X_c \tau \nu$)
decay is sensitive to new physics,in particular, models with charged Higgs 
bosons.Because the charged Higgs bosons contribute at tree level,its 
contribution can not be cancelled by other new particles in the models.
Therefore,the calculations of charged Higgs contributions with high accuracy
will provide strong bound on parameters of the models when experimental 
measurement of the decay is available.

In this paper we investigate the polarized inclusive decay $\Lambda_b \rightarrow 
X_c \tau \nu$ in both the SM and THDM. In the SM we extend the results of
 Manohar and Wise\cite{3} to the case of non-zero mass of the final state lepton
(tau). We calculate the spin dependent form factors in the hadronic tensor to the 
1/$m^2_b$
 order in HQET which do not contribute to the decay rate when the mass of the
final state lepton is neglected. In the THDM we compute the inclusive rate and
$\tau$ spectrum for a polarized $\Lambda\rightarrow X_c\tau\nu$ included the $
\Lambda^2_{QCD}$/$m^2_b$ nonperturbative corrections and the $O(\alpha_s)$
perturbative corrections to $\tau$ spectrum.
 
 The paper is organized as follows. In section II we calculate $\tau$ 
spectrum from a polarized $\Lambda_b$ decay to the order $1/m_b^2$ in the  $1/m_b$
expansion in the SM.
Section III is devoted to calculations in the THDM. The $O(\alpha_s)$ QCD corrections
to the decay are included. In section IV numerical
results are given. Finally, a summary and discussions are presented in section
V.

\begin{center}
{\large \bf II.$\tau$ spectrum of $\Lambda_b \rightarrow X_c \tau \nu$ in the SM}
\end{center}

We consider the inclusive semileptonic decay $\Lambda_b \rightarrow X_c 
\tau \nu$ in the SM. At the tree level,for unpolarized leptons the partial
decay width can be written as
\begin{equation}
d\Gamma=\frac {G_f^2 |V_{cb}|^2}{(2\pi)^5 E_H} L^{\mu\nu}W_{\mu\nu}
\frac {d^3 p_\tau}{2 E_\tau} \frac {d^3 p_\nu}{2 E_\nu},            
\end{equation}
where $E_H$ is the energy of $\Lambda_b$,$L_{\mu\nu}$ is the leptonic tensor
\begin{equation}
L^{\mu\nu}=2(p_\tau^\mu p_\nu^\nu+p_\tau^\nu p_\nu^\mu-g^{\mu\nu} p_\tau.
p_\nu+i\epsilon^{\mu\nu\alpha\beta} (p_\tau)_\alpha (p_\nu)_\beta)  ,   
\end{equation}
and $W_{\mu\nu}$ is the hadronic tensor
\begin{equation}
W_{\mu\nu}=(2\pi)^3 \sum\limits_{X} \delta^4 (p_{\Lambda_b}-q-p_X) \langle
\Lambda_b
(v,s)|{J_\mu}^{\dagger}|X\rangle \langle X|J_\nu|\Lambda_b(v,s) \rangle   
\end{equation}
with $p_{\Lambda_b}=m_{\Lambda_b} v, q=p_\tau+p_\nu$ and s being the spin of 
$\Lambda_b$.
$J_\mu=\bar{c} \gamma_\mu \frac {(1-\gamma_5)}{2} b$ in eq.(3) is the hadronic current.
The expansion of $W_{\mu\nu}$ in terms of Lorentz invariant structure functions
$W_i$ is defined by
\begin{eqnarray}
W_{\mu\nu} &=& -g_{\mu\nu} W_1+v_\mu v_\nu W_2-
i \epsilon_{\mu\nu\alpha\beta} v^\alpha q^\beta W_3+ 
q_\mu q_\nu W_4+(q_\mu v_\nu+q_\nu v_\mu)W_5+ \left\{ - q.s \right.  \cr
& & \left. \left[-{\it g_{\mu\nu}} W_1^s+{\it v_\mu v_\nu} W_2^s-
{\it i \epsilon_{\mu\nu\alpha\beta}
v^\alpha q^\beta} W_3^s+{\it q_\mu q_\nu} W_4^s+\left({\it q_\mu v_\nu}+
{\it q_\nu v_\mu} \right) W_5^s \right] +  \right. \cr
& & \left.\left({\it s_\mu v_\nu}+{\it s_\nu v_\mu }\right) W_6^s+
\left({\it s_\mu q_\nu}+{\it s_\nu q_\mu} \right) W_7^s+
{\it i \epsilon_{\mu\nu\alpha\beta} v^\alpha s^\beta} W_8^s+
{\it i \epsilon_{\mu\nu\alpha\beta} q^\alpha s^\beta} W_9^s  \right\}. 
\end{eqnarray}
The structure functions $W_i$ can be calculated in heavy quark effective
theory (HQET)\cite{15} and results to the $1/m_b^2$ order are
\begin{eqnarray}
W_1 &=& {\it \delta(z)}\,\left( {{{\it m_b}}\over 2} + {{K_b\,{\it m_b}}\over 6} - 
      {{{\it q.v}}\over 2} \right)  - 
   {{{\it \delta^{'}(z)}}}\,\left( {{2\,K_b\,{\it m_b}\,{q^2}}\over 3} + 
      K_b\,{{{\it m_b}}^2}\,{\it q.v} - {{5\,K_b\,{\it m_b}\,{{{\it q.v}}^2}}\over 3}
       \right)  + \cr
& &       {{{\it \delta^{''}(z)}}}\,
    \left( {{-2\,K_b\,{{{\it m_b}}^3}\,\left({q^2}-q.v^2 \right)}\over 3} + 
      {{2\,K_b\,{{{\it m_b}}^2}\,{\it q.v}\,\left({q^2}-q.v^2 \right)}\over 3}  
       \right) ,\cr
W_2 &=& {\it \delta(z)}\,\left( {\it m_b} + {{5\,K_b\,{\it m_b}}\over 3} \right)  -
   {{14\,{{{\it \delta^{'}(z)}}}\,K_b\,{{{\it m_b}}^2}\,{\it q.v}}\over 3} + 
   {{{\it \delta^{''}(z)}}}\,\left( {{-4\,K_b\,{{{\it m_b}}^3}\,
\left({q^2}-q.v^2 \right)}\over 3} \right) , \cr
W_3 &=& {{{\it \delta(z)}}\over 2} - 
  {{5\,{{{\it \delta^{'}(z)}}}\,K_b\,{\it m_b}\,{\it q.v}}\over 
     3} + {{{\it \delta^{''}(z)}}}\,\left( {{-2\,K_b\,{{{\it m_b}}^2}\,
\left({q^2}-q.v^2 \right)}\over 3} \right)  ,\cr
W_4 &=& {{-4\,{{{\it \delta^{'}(z)}}}\,K_b\,{\it m_b}}\over 3} ,\cr
W_5 &=& {{-{\it \delta(z)}}\over 2} - {{{\it \delta^{'}(z)}}}\,
    \left( {{-4\,K_b\,{{{\it m_b}}^2}}\over 3} - 
      {{5\,K_b\,{\it m_b}\,{\it q.v}}\over 3} \right)  + 
   {{{\it \delta^{''}(z)}}}\,\left( {{2\,K_b\,{{{\it m_b}}^2}\,
\left({q^2}-q.v^2 \right)}\over 3} \right) ,\cr
W_1^s &=& {{- {\it \delta(z)}\,\left( 1 + {\it \epsilon_b} \right) }\over 2} -
  {{{\it \delta}}^{'}(z)}\, 
      {{5\,K_b\,{\it m_b}\,\left( - {\it q.v} \right) }\over 3} + 
      {{2\,{{{\it \delta}}^{''}(z)}\,K_b\,{{{\it m_b}}^2}\,
       \left( {q^2} - {{{\it q.v}}^2} \right) }\over 3} ,\cr
W_2^s &=& {{-4\,{{{\it \delta}}^{'}(z)}\,K_b\,{{{\it m_b}}^2}}\over 3} ,\cr
W_3^s &=& {{-2\,{{{\it \delta}}^{'}(z)}\,K_b\,{\it m_b}}\over 3} ,\cr
W_4^s &=& 0,\cr
W_5^s &=& {{2\,{{{\it \delta}}^{'}(z)}\,K_b\,{\it m_b}}\over 3} ,\cr
W_6^s &=& {\it \delta(z)}\,\left( {{- \left( 1 + {\it \epsilon_b} \right) \,
            {\it m_b} }\over 2} - {{5\,K_b\,{\it m_b}}\over 6} \right)  -
   {{{\it \delta}}^{'}(z)}\,\left( -{{5\,K_b\,
   {{{\it m_b}}^2}\,{\it q.v}}\over 3} \right)  + 
 {{2\,{{{\it \delta}}^{''}(z)}\,K_b\,{{{\it m_b}}^3}\,
       \left( {q^2} - {{{\it q.v}}^2} \right) }\over 3} ,\cr
W_7^s &=& {{{\it \delta(z)}\,\left( 1 + {\it \epsilon_b} \right) }\over 2} -
   {{{\it \delta}}^{'}(z)}\,\left( {{5\,K_b\,{{{\it m_b}}^2}}\over 3} - 
      K_b\,{\it m_b}\,\left( {\it m_b} - {\it q.v} \right)  \right)  - \cr
& &   {{2\,{{{\it \delta}}^{''}(z)}\,K_b\,{{{\it m_b}}^2}\,
       \left( {q^2} - {{{\it q.v}}^2} \right) }\over 3} ,\cr
W_8^s &=& {\it \delta(z)}\,\left( {{\left( 1 + 
{\it \epsilon_b} \right) \,{\it m_b}}\over 2} + 
      {{K_b\,{\it m_b}}\over 6} \right)  -
  {{{\it \delta}}^{'}(z)}\,\left({{5\,K_b\,{{{\it m_b}}^2}\,{\it q.v}}\over 3} \right)  - 
   {{2\,{{{\it \delta}}^{''}(z)}\,K_b\,{{{\it m_b}}^3}\,
       \left( {q^2} - {{{\it q.v}}^2} \right) }\over 3} ,\cr
W_9^s &=& {{-{\it \delta(z)}\,\left( 1 + {\it \epsilon_b} \right) }\over 2} -
   {{{\it \delta}}^{'}(z)}\,\left( {{-5\,K_b\,{{{\it m_b}}^2}}\over 3} + 
      K_b\,{\it m_b}\,\left( {\it m_b} - {\it q.v} \right)  \right)  + \cr
& &   {{2\,{{{\it \delta}}^{''}(z)}\,K_b\,{{{\it m_b}}^2}\,
       \left( {q^2} - {{{\it q.v}}^2} \right) }\over 3} .
\end{eqnarray}
where 
\begin{equation}
K_b=-\langle \Lambda_b(v,s)|\bar b_v {\frac {(iD)^2}{2m_b^{2}}} b_v|
\Lambda_b(v,s)\rangle, z=(m_b v-q)^2-m_c^2 ,
\end{equation}
and $\epsilon_b$ is defined by\cite{3}
\begin{equation}
\langle \Lambda_b(v,s)|\bar b \gamma^\lambda \gamma_5 b|\Lambda_b(v,s)
\rangle=(1+\epsilon_b) \bar{u}(v,s) \gamma^\lambda \gamma_5 u(v,s).
\end{equation}
$K_b$ and $\epsilon_b$ are only unknown two parameters at $O(m_b^{-2})$ for
the $\Lambda_b$ decay which parametrize the nonperturbative phenomena and 
are expected to be of order $(\Lambda_{QCD}/m_b)^2$.We will discuss them in 
section  IV.The another parameter at $O(m_b^{-2})$ 
\begin{equation}
G_b=Z_b \langle H_b(v,s)|\bar b_v {\frac {g G_{\alpha \beta} \sigma^{\alpha \beta}}
{4m_b^{2}}} b_v|H_b(v,s)\rangle 
\end{equation}
is equal to zero for $H_b=\Lambda_b$ due to the zero spin of the light degrees 
of freedom inside $\Lambda_b$. $W_i$(i=1,2,3) and $W_i^s$(i=1,2,3,6,8,9) have already
been given by Manohar and Wise\cite{3} and $W_i$(i=4,5) by Balk et al.\cite{5}. We list them here
only for completeness.
>From eqs(1),(2),(4) and(5) we get the differential decay rate
\begin{equation}
\frac {d\Gamma_W}{\Gamma_b dtdxdydcos\theta}=\hat A (x,t,y,\eta,\epsilon)+
   \hat B (x,t,y,\eta,\epsilon) cos\theta,
\end{equation}
here
\begin{eqnarray}
\hat A(x,t,y,\eta,\epsilon)&=&[ -12\,t\,y + 12\,x\,y - 12\,y\,{\it \eta} +
     K_b\,\left( -16\,t + 20\,x\,y + 16\,{\it \eta} \right)  ] \,
   {\it \delta}(z) - \cr
& &   4\,K_b\,\left( 4\,{t^2} - 4\,t\,x - 4\,t\,y - 5\,t\,x\,y +
     7\,{x^2}\,y - 5\,t\,{y^2} + 7\,x\,{y^2} + 4\,x\,{\it \eta} \right. \cr
& & \left.    - 4\,y\,{\it \eta} - 5\,x\,y\,{\it \eta} - 5\,{y^2}\,{\it \eta} -
     4\,{{{\it \eta}}^2} \right) \,{\it \delta}'(z) +
  4\,K_b\,y\,\left( 4\,t - {x^2} - 2\,x\,y  \right. \cr
& & \left.  - {y^2} \right) \,
   \left( t - x + {\it \eta} \right) \,{\it \delta}''(z) ,\cr
\hat B(x,t,y,\eta,\epsilon)&=&{{1}\over{\sqrt{x^2-4 \eta}}} \{[  
24\,{t^2} - 
   24\,t\,x - 12\,t\,x\,y + 12\,{x^2}\,y + 24\,x\,{\it \eta} -
     12\,x\,y\,{\it \eta} -   \cr
& &    24\,{{{\it \eta}}^2} +
     K_b\,\left( -24\,t\,x + 20\,{x^2}\,y + 24\,x\,{\it \eta} -
        32\,y\,{\it \eta} \right)  +
     {\it \epsilon_b}\,\left( 24\,{t^2} - 24\,t\,x  \right. \cr
& &  \left. - 12\,t\,x\,y + 12\,{x^2}\,y +
        24\,x\,{\it \eta} - 12\,x\,y\,{\it \eta} - 24\,{{{\it \eta}}^2} \right)
      ] \,{\it \delta}(z) - 4\,K_b\,
   \left( 8\,{t^2} +\right. \cr
& & \left.  6\,{t^2}\,x - 10\,t\,{x^2} + 10\,{t^2}\,y -
     18\,t\,x\,y - 5\,t\,{x^2}\,y + 7\,{x^3}\,y - 5\,t\,x\,{y^2} +  \right.\cr
& & \left.  7\,{x^2}\,{y^2} + 10\,{x^2}\,{\it \eta} + 8\,t\,y\,{\it \eta} +
     2\,x\,y\,{\it \eta} - 5\,{x^2}\,y\,{\it \eta} - 5\,x\,{y^2}\,{\it \eta} -
     8\,{{{\it \eta}}^2} -  \right. \cr
& &   \left.  6\,x\,{{{\it \eta}}^2} - 2\,y\,{{{\it \eta}}^2} \right)
    \,{\it \delta}'(z) + 4\,K_b\,\left( 4\,t - {x^2} - 2\,x\,y - {y^2} \right) \,
   \left( t - x + {\it \eta} \right) \,
   \left( \right. \cr
& &  \left. -2\,t + x\,y + 2\,{\it \eta} \right) \,{\it \delta}''(z) \}
\end{eqnarray}
where 
$$\Gamma_b=\frac {G_F^2 |V_{cb}|^2 m_b^5}{192\pi^3},
x=\frac {2E_\tau}{m_b},y=\frac {2E_\nu}{m_b},t=\frac {q^2}{m_b^2},
\eta=\frac {m_\tau^2}{m_b^2},\epsilon=\frac {m_c}{m_b}$$
and $\theta$ is the angle between the $\tau$ direction and the $\Lambda_b$ spin
in the rest frame of the $\Lambda_b$.
After integrating over t and y,one obtains the $\tau$ energy spectrum
\begin{equation}
\frac {d\Gamma_W}{\Gamma_b dxdcos\theta}=A_W(x,\eta,\epsilon)+B_W(x,\eta,\epsilon) cos\theta,
\end{equation}
where $A_W(x,\eta,\epsilon)$ and $B_W(x,\eta,\epsilon)$ are given in Appendix.
The $\eta\rightarrow 0$ limits of $A_W$ and $B_W$ agree with the results of Manohar 
and Wise\cite{3}. The total inclusive decay width of $\Lambda_b \rightarrow 
X_c \tau \nu$ can be obtained by integrating the spectrum formula over range
$$2 \sqrt{\eta}\leq x \leq 1-\rho+\eta,$$
the result is not present here because one can easily obtain it from ref.\cite{5} 
by taking $G_b$=0. 
Perturbative O($\alpha_s$) QCD corrections to the double differential distribution of 
the $\tau$ energy and the invariant mass of the lepton system for $b\rightarrow c\tau 
\nu$ have been studied by M. Je\.zabek et al.\cite{16,17}.
We will use eq.(30) in ref.\cite{17}
in our numerical analysis for the nonpolarized distribution of the $\tau$ energy.
\begin{center}
{\large \bf III.$\tau$ spectrum of $\Lambda_b \rightarrow X_c \tau \nu$ in THDM}
\end{center}

We consider the THDM\cite{18,19,20} in which the up-type quarks get masses from 
Yukawa
couplings to the one Higgs doublet $H_2$(with the vacuum expectation value
$v_2$) and down-type quarks and leptons get masses from Yukawa couplings to
the another Higgs doublet $H_1$ (with the vacuum expectation value $v_1$). Such
a model occurs as a natural feature in supersymmetric theories. For the sake of
simplicity we shall use the Feynman rules of the THDM in MSSM\cite{21}.In a 
THDM there
are three diagrams contributing to the decay $\tau$ spectrum of $\Lambda_b 
\rightarrow X_c \tau \nu$ which correspond to W-exchange,Goldstone 
boson-exchange,and charged Higgs boson-exchange respectively if one uses a
non-physical gauge. It is shown in ref.\cite{8} that in the Landau gauge the rate can
be decomposed into the sum of two incoherent decays:
$$M=M_W+M_S \Rightarrow |M|^2=|M_W|^2+|M_S|^2$$
where $M_S=M_G+M_H$.$M_W,M_G$ and $M_H$ are the W-mediated,Goldstone
boson-mediated and Higgs boson-mediated decay amplitudes respectively.
This decomposition has an advantage to simplify calculations,in particular,
the calculations of QCD corrections. We assume the Landau gauge hereafter.Then 
the new thing needed to do is to calculate $M_S$.
 
For the purpose of calculating $M_S$ the hadronic current $J_\mu$ in eq.(3) is replaced by
\begin{equation}
J_i=\bar {c} (a_i+b_i \gamma_5) b \,\,\,\, (i=H,G)
\end{equation}
with
\begin{eqnarray}
\nonumber a_H=m_b tan\beta+m_c cot\beta,b_H=m_b tan\beta-m_c cot\beta,a_G=-m_b+m_c,b_G=-m_b-m_c.
\end{eqnarray}
Following the same steps as those in section II,a straightforward calculation
leads to
\begin{eqnarray}
\frac {d\Gamma_H}{\Gamma_b dxdcos\theta}&=&A_H(x,\eta,\epsilon,\xi,tan\beta)+
B_H(x,\eta,\epsilon,\xi,tan\beta)cos\theta,\\
\frac {d\Gamma_I}{\Gamma_b dxdcos\theta}&=&A_I(x,\eta,\epsilon,\xi,tan\beta)+
B_I(x,\eta,\epsilon,\xi,tan\beta) cos\theta,
\end{eqnarray}
with $A_i(x,\eta,\epsilon,\xi,tan\beta)$ and $B_i(x,\eta,\epsilon,\xi,tan\beta) (i=H,I)$given in Appendix.
Here $d\Gamma_i/dx (i=H,I)$ denotes the contributions to the $\tau$
spectrum from Higgs-mediated and the interference term between Higgs-mediated
and Goldstone boson-mediated respectively,and $\xi=m_H/m_b$.

For the spin independent terms $A_H$ and $A_I$, our results agree with those
obtained by Y. Grossman et al.\cite{7}.
Note that $B_I=0$ at the leading 
order of $1/m_b$ expansion as $m_c \rightarrow 0$.This is due to the chiral
difference of the vertices of W and Higgs. As we can see in numerical
analysis
that this makes $B_I$ much smaller than $B_W$.

Combining eq.(11),eq.(13) and eq.(14), one obtains the $\tau$ spectrum of
$\Lambda_b \rightarrow X_c \tau \nu$ in the THDM
\begin{equation}
\frac {d\Gamma_{THDM}}{\Gamma_b dxdcos\theta}=(A_W+A_I+A_H)+(B_W+B_I+B_H)cos\theta.
\end{equation}
We now come to the position to calculate $O(\alpha_s)$ QCD corrections.
Making use of the results obtained by A.Czarnecki et al.\cite{22},Y. Grossman 
et al.
got the $O(\alpha_s)$ corrections of the total width of $b \rightarrow c \tau \nu$ 
mediated by Higgs\cite{8}.
To get the $O(\alpha_s)$ corrections of $\tau$ spectrum one can also use 
their results. We find that the 
relation between $d\Gamma_{\alpha_s}^{H(I)}/dxdt$ and 
$d\Gamma_{\alpha_s}^{H(I)}/dt$ is very simple, as expected.
$d\Gamma_{\alpha_s}^{H(I)}/dxdt$ is independent of x because Goldstone
boson and Higgs are both scalar particles. 
Therefore,$d\Gamma_{\alpha_s}^{H(I)}/dxdt$ can be
 simply obtained by dividing $d\Gamma_{\alpha_s}^{H(I)}/dt$ by $x_{max}-x_{min}$,
where $x_{max},x_{min}$ denote $x$'s kinematical upper and 
lower limits respectively.
$d\Gamma_{\alpha_s}^{H(I)}/dt$ can be easily obtained from eq.(8) of ref.\cite{22}
by multiplying the lepton part.The results are
\begin{eqnarray}
& & \frac {d\Gamma_{\alpha_s}^{H}}{dxdt}=\frac {\sqrt {2} m_b^2 \eta (t-\eta) 
tan^2\beta}{16 \pi^2 \xi^4 p_3}\Gamma (c_H^1,c_H^2,c_H^3),\cr
& & \frac {d\Gamma_{\alpha_s}^{I}}{dxdt}=\frac {\sqrt {2} m_b^2 \eta (t-\eta) 
tan\beta}{8 \pi^2 \xi^2 t p_3}\Gamma (c_I^1,c_I^2,c_I^3),
\end{eqnarray}
here 
\begin{eqnarray}
& & c_H^1=2 tan^2\beta+2 \epsilon^2 cot^2\beta,c_H^2=4 \epsilon,
c_H^3=tan^2\beta-\epsilon^2 cot^2\beta,\cr
& & c_I^1=-2 tan\beta+2 \epsilon^2 cot\beta,c_I^2=2 \epsilon (tan\beta-cot\beta),
c_I^3=-tan\beta-\epsilon^2 cot\beta, 
\end{eqnarray}
and $\Gamma (c^1,c^2,c^3)$ is\cite{22} 
\begin{equation}
\Gamma(c^1,c^2,c^3)=\frac {\alpha_s}{6 \pi^2} \frac {G_F m_b^3 |V_{cb}|^2}{\sqrt{2}}
  \left[ c^1 G_{+}+c^2 G_{-}+c^3 G_0 \right]
\end{equation}
with
\begin{eqnarray*}
G_{+}&=&p_0 H+p_0 p_3 [ \frac {9}{2}-2\ln(\frac {16 p_3^4}{\epsilon^2 t})]+
  \frac {1}{4 t} Y_p (2-t-4t^2+3 t^3-2 \epsilon^2-2 \epsilon^4+
  2 \epsilon^6-4 t \epsilon^2-5 t \epsilon^4) ,\cr
G_{-}&=&H+p_3 [6-2 \ln(\frac{16 p_3^4}{\epsilon^2 t})]+\frac{1}{t} Y_p (1-t-
  2 \epsilon^2+\epsilon^4-3 t \epsilon^2),\cr
G_0&=&-6 p_0 p_3 \ln\epsilon,
\end{eqnarray*}
where
$$H=4p_0 [Li_2 (p_{+})-Li_2 (p_{-})-2Li_2 (1-\frac{p_{-}}{p_{+}})+\frac{1}{2}
   Y_p \ln(\frac {16 p_3^4 t}{p_{+}^4}) -Y_w \ln\epsilon ]+2 Y_w (1-\epsilon^2)+
   \frac{2}{t} p_3 \ln\epsilon (1+t-\epsilon^2),$$
\begin{eqnarray*}
&&  p_0 \equiv \frac{1}{2} (1-t+\epsilon^2),
  p_3 \equiv \frac{1}{2} \sqrt {1+t^2+\epsilon^4-2(t+\epsilon^2+t \epsilon^2)},
  p_{\pm} \equiv p_0 \pm p_3, \cr
&&  Y_p \equiv \frac{1}{2} \ln \frac{p_{+}}{P_{-}},W_0 \equiv \frac {1}{2} (1+t-\epsilon^2),
 W_{\pm} \equiv W_0 \pm p_3 , Y_w \equiv \frac {1}{2} \ln \frac{W_{+}}{W_{-}}.
\end{eqnarray*}
After integrating (16) over t, we get $\tau$ spectrum numerically with only 
parameters $tan\beta$ and $\xi$.
As a check,we find that our numerical results of $\Gamma_{\alpha_s}^H$ and
$\Gamma_{\alpha_s}^I$ agree with those obtained by Y. Grossman et al.\cite{8}.

\begin{center}
{\large \bf IV. Numerical Results}
\end{center} 
In order to do numerical calculations we need to discuss values of parameters which are 
$K_b, \epsilon_b, m_c, m_b$ and the parameters in THDM.

1.Constraints on the parameters of THDM

In ref.\cite{23,24},constraints on tan$\beta$ from $K-\bar {K}$ and $B-\bar 
{B}$ mixing
,$\Gamma(b\rightarrow s \gamma),\Gamma(b\rightarrow c \tau \nu)$ and $R_b$ have
been given
$$0.7 <tan\beta <0.6 \frac {m_{H_\pm}}{1 Gev}$$
and also the lower limit ${m_{H_\pm}}>200 Gev$ has been given there. Taking 
the radiative correction and the $1/m_Q^2$ correction into account in
the B meson decay, Y.Grossman et al. have a improved bound of R (defined by 
$R=tan\beta/m_{H^{\pm}}$) which is $$R<0.49 Gev^{-1}$$.
We will predict the $\tau$ spectrum and total width of $\Lambda_b$ decay under
these constraints.

2.About the parameters $K_b,\epsilon_b$ and $m_b,m_c$

$K_b$ and $\epsilon_b$ characterize the 1/$m_Q^2$ corrections to the decay distribution 
for $\Lambda_b\rightarrow X_c\tau\nu$ and are nonperturbative quantities independent of $m_Q$.
Since quarks are not free physical particles,$m_b$ and $m_c$ can not be
determined directly by experiment. However,
we can estimate them by the phenomenological analysis of the heavy hadron spectra to the order 1/$m_Q$. 
>From the effective Lagrangian in HQET, the mass of a heavy hadron can 
be written as\cite{28,29,30}
\begin{equation}
m_{h_Q}=m_Q+\bar {\Lambda}(j^P_l, I, S)+\frac {a(j^P_l, I, S)}{ m_Q}+\frac{b(j^P_l, I, S)}{m_Q} \langle\vec{S}_Q\cdot\vec{j}_l\rangle+...,
\end{equation}
where $\langle\vec{S}_Q\cdot\vec{j}_1\rangle=\frac{1}{2}[J(J+1)-j_l(j_l+1)-\frac{3}{4}]$ with J and $j_l$ 
being the spins of the hadron and the light degrees of freedom inside the hadron respectively. 
The parameter $\bar {\Lambda}$ represents 
contributions come from the effective Lagrangian in the $m_Q\rightarrow\infty$ limit,
 and a, b are
respectively associated with the kinetic energy and the color magnetic energy of the 
heavy quark inside the hadron. In present case,$a(0^+, 0, 0)= m_b^2 K_b$ and
b$(0^+,0,0)$=0. It is shown that $\epsilon_b\leq -2/3 K_b$\cite{27}. 
Furthermore, one can take 
$\epsilon_b = -2/3 K_b$ if one omits the contributions of terms arising from double
insertions of chromomagnetic operator\cite{25,26,27}. Starting from eq.(19), it 
is shown\cite{30} 
that one can obtain $a(\frac{1}{2}^-, \frac{1}{2}, 0)$, $m_c$ (and $\bar{\Lambda}
(\frac{1}{2}^-, \frac{1}{2}, 0), b(\frac{1}{2}^-, \frac{1}{2}, 0)$ ) by using the 
observed masses of the doublets $(B^{*}, B)$ and $(D^{*}, D)$ if choosing $m_b$ as input.
 Furthermore, $a(0^+, 0, 0)$ which is the parameter we need is determined 
by\cite{30,3} 
\begin{equation}
a(0^+, 0, 
0)=\frac{h\,m_b}{1-h}[(m_{\Lambda_c}-\bar{m}_D)-(m_{\Lambda_b}-\bar{m}_B)]+a(\frac{1}{2}^-, \frac{1}{2}, 0), \nonumber
\end{equation}
where
\begin{equation}
h=\frac{m_{B^*}-m_B}{m_{D^*}-m_D}, \nonumber
\end{equation}
and $\bar{m}_H=\frac{1}{4}(m_H+3m_{H^*})$, H=B, D. 
Therefore, if we choose $m_b=5.1$Gev,
the other parameters will be$\dagger$
\footnotetext{$\dagger$ the experimental data used in ref.\cite{30} have been 
improved since then. Our data are from ref.\cite{31}.}:
$$m_c=h\,m_b=1.65Gev$$
$$K_b=\frac {0.142 m_c Gev}{m_b^2}\approx 0.009,\epsilon_b \approx-0.006$$

If we choose $m_b=5.044$ which is a "critical" value based on eq.(19)\cite{30}, 
other parameters will be
$$m_c=1.63Gev,K_b\approx 0.006,\epsilon_b \approx -0.004$$

We will use these two sets of values and discriminate them by the 
first($m_b=5.044 Gev$) 
and second($m_b=5.1 Gev$) set respectively in the numerical computations.

Using the parameters given above, we obtain the total width in terms of tan$\beta$ and $m_H$ as follows:
\begin{eqnarray}
& & \frac{\Gamma_W^1}{\Gamma_b}=C_W^1+D_W^1 \alpha_s,\cr
& &\frac{\Gamma_H^1+\Gamma_I^1}{\Gamma_b}=C_H^1+D_H^1 \alpha_s,,\cr
& &\frac{\Gamma_W^2}{\Gamma_b}=C_W^2+D_W^2 \alpha_s,\cr
& &\frac{\Gamma_H^2+\Gamma_I^2}{\Gamma_b}=C_H^2+D_H^2 \alpha_s,
\end{eqnarray}
where 
\begin{eqnarray}
C_W^1&=&0.109,\cr
C_H^1&=&-{{0.0141\,\left( 0.253 + 1.16\,{{{\it tan}}^2 \beta} \right) }\over
    {{{{\it \xi}}^2}}} + {{0.0141\,
      \left( 0.025 + 0.112\,{{{\it tan}}^2 \beta} +
         0.239  \,{{{\it tan}}^4 \beta} \right) }\over
    {{{{\it \xi}}^4}}},\cr
D_W^1&=&-0.0476,\cr
D_H^1&=&{{0.00804\,\left( -0.165 - 0.577\,{{{\it \xi}}^2} - 0.374\,{{{\it 
tan}}^2\beta} -
       {{{\it \xi}}^2}\,{{{\it tan}}^2\beta} - 0.331\,{{{\it tan}}^4\beta} \right) }
    \over {{{{\it \xi}}^4}}},\cr
C_W^2&=&0.112,\cr
C_H^2&=&-{{0.0139\,\left( 0.262 + 1.19\,{{{\it tan}}^2\beta} \right) }\over
    {{{{\it \xi}}^2}}} + {{0.0139\,
      \left( 0.0256 + 0.107\,{{{\it tan}}^2\beta} +
         0.245  \,{{{\it tan}}^4\beta} \right) }\over
    {{{{\it \xi}}^4}}},\cr
D_W^2&=&-0.0493, \cr
D_H^2&=&{{0.0082\,\left( -0.163 - 0.574\,{{{\it \xi}}^2} - 0.37\,{{{\it 
tan}}^2\beta} -
       {{{\it \xi}}^2}\,{{{\it tan}}^2\beta} - 0.328\,{{{\it tan}}^4\beta} \right) }
    \over {{{{\it \xi}}^4}}},
\end{eqnarray}
The superscript i (i=1, 2) in eqs.(22) and (23) denotes that the ith set of values 
of $K_b, \epsilon_b, m_c$ and $m_b$ is used.

The explicit dependence of total width $\Gamma_{THDM}=\Gamma_W+\Gamma_H+\Gamma_I$ 
(normalized to the electron channel) on tan$\beta$ and $m_H$ is plotted in 
fig.1. It should be noted that 
the absolute value of $\Gamma_{THDM}$ is very sensitive to $m_b$. Using the ratio 
between the width $\Gamma_{THDM}^{\tau}$ and $\Gamma_{THDM}^e$ separates
the theoretical
and experimental uncertainties and deletes the $|V_{cb}|^2 m_b^5$ factor which
is not well-known yet. From fig.1 the following remarks can be drawn:
(1) The normalized total width is not sensitive to the values of the 
parameters in 
HQET. The value of the normalized width for the second set of parameters is 
three percent
larger than that for the first set. (2) In the range of tan$\beta$ which is interesting
physically, say, tan$\beta<60$, the normalized width changes roughly five to 
fifteen percent when
tan$\beta$ changes from twenty to sixty and $m_{H^{\pm}}$ is fixed. The 
normalized width changes
the same order of magnitude for fixed tan$\beta$ and changing $m_{H^{\pm}}$ from 
200 to 
400 Gev. From eqs. (22) and (23) it follows that the $\alpha_s$ corrections 
decrease the total width roughly 20 percent which
is larger than that in SM. Because when tan$\beta\gg$ 1 $C_H^i$ is
proportional to $r^2\,(-1+0.2\, r^2)$ and $D_H^i$ is proportional to $-r^2\,
(1+0.3\,r^2)$
where r=$R\,m_b$, one can obtain constraints on $r$ from the measurement
of the total width.  

$\tau$ spectrum for some typical values of tan$\beta$ and $m_H$ is calculated
and the result 
is plotted in figs.2-5. The predictions in the SM are also plotted in the 
figs.. Here the 
$\alpha_s$ corrections for 
the spin-dependent term are not considered. We can see from figs.4-5 that the spin-dependent
spectrum is quite different for $r\leq 1$ and $r \geq 2$ (Note that we have the constraints
$r<0.49\,Gev^{-1}\,m_b\approx 2.5$ and $m_{H^{\pm}}\geq 200 Gev$ from experiments, as 
mentioned before). The reason is as 
follows. We know from Appendix that $B_H=-\frac{J\,r^2\,\eta}{8\,\xi^2}\,B_W\approx -r^4\,
\eta\,B_W/4$. That is, it depends $r^4$. 
For $r\leq 1$, $B_H\ll\,B_W$. 
As pointed in section III, $B_I$ is negligiablly
small comparing with $B_W$
due to the chiral difference of b quark couplings to 
W and $H_\pm$ which is deduced from Model II of THDM. Therefore, the spectrum is 
almost the same as that in the SM, as can be seen from figs.4-5. For $r\geq 2$
, $B_H$ is as the same order of magnitude as $B_W$
so that $B_H$ and $B_W$ tend to cancel each other, which 
makes the spin-dependent distribution of $\tau$ energy very small and a little
dependent of the $\tau$ energy, as shown in figs.4-5. Thus one can say that if the 
$\tau$ spectrum is somewhat more isotropic than what the SM predicts, the THDM with large
tan$\beta$ ($>80$) and $m_H\geq 200Gev$ is preferred in describing the nature.
For the nonpolarized term, as can be seen from figs.4-5, the spectrum is very 
similar to that of B decay since the difference between the $\Lambda_b$
decay and B decay comes from the 1/$m_b^2$ corrections. 

\begin{center}
{\large \bf V. Summary}
\end{center}

In summary, we have calculated the rate and $\tau$ spectrum of the inclusive semileptonic
decay for a polarized $\Lambda_b\rightarrow X_c\tau\nu$ to the 1/$m_b^2$ order in the
1/$m_b$ expansion in the SM. The $\alpha_s$ corrections are included in the numerical 
computations for the spin independent terms of $\tau$ spectrum. Our results show that
the spin dependent $\tau$ spectrum is significant enough to be seen.

We have also calculated the same quantities in a THDM. For the spin independent terms
of $\tau$ spectrum arising from Higgs-mediated and the interference term, we have 
calculated the $O(\alpha_s)$ QCD corrections to the double differential distribution. 
Together with the $\alpha_s$ corrections
in the SM given in ref.\cite{16}, we obtained all the $\alpha_s$ corrections 
to the 
non-polarized double differential distribution (and so the total width) in the THDM. 
The numerical results show that the branching ratio of 
$\Lambda_b\rightarrow X_c\tau\nu$ in THDM is of approximately 25 percent 
of that in the electron channel and the spin dependent $\tau$ spectrum can be used
to estimate the size of tan$\beta$ and $m_{H^{\pm}}$. 
The spectrum depends dominantly on R if tan$\beta\gg 1$ so that from 
the measurement of the angular distribution of a polarized 
$\Lambda_b\rightarrow X_c\tau\nu$ in B-factories within the coming years one can 
obtain constraints on R.  

It is obvious that substituting the u quark mass $m_u$=o for $m_c$ one immediately
obtains the decay rate and $\tau$ spectrum for a polarized $\Lambda_b\rightarrow X_u
\tau\nu$. And with minor changes one can extend the results in the paper 
to the
inclusive semileptonic decay of a polarized $\Lambda_c\rightarrow X_{s,d}
\tau\nu$. It is interesting to calculate the $\alpha_s$ corrections 
to the spin dependent term of $\tau$ spectrum. 

{\large \bf Acknowledgments}

One of the authors(H-G Yan) thanks Y. Grossman and F. Kr\"uger for 
enlightening 
discussions. The other (C.-S. Huang) wishes to acknowledge International Centre for 
Theoretical Physics where the paper was written 
for the warm hospitality. We thank Y.P. Yao for kindly offering his 
unpublished article,
L. Motyka and M. Je\.zabek for their program concerning the formulae of 
ref.\cite{17}.

\vspace*{1.5cm}

\begin{center}
{\large \bf Appendix: Expressions of $A_i$ and $B_i(i=W,H,I)$ }
\end{center}
\begin{eqnarray}
A_W &=& {{{\sqrt{{x^2} - 4\,{\it \eta}}}}\over 
   {{{\left( 1 - x + {\it \eta} \right) }^3}}}\,
     {{\left( -1 + {\it \epsilon^2} + x - {\it \eta} \right) }^2}\,
     \left( 3\,x + 3\,{\it \epsilon^2}\,x - 5\,{x^2} - {\it \epsilon^2}\,{x^2} + 2\,{x^3} - \right. \cr
& & \left.   4\,{\it \eta} - 8\,{\it \epsilon^2}\,{\it \eta} + 10\,x\,{\it \eta} + 
       3\,{\it \epsilon^2}\,x\,{\it \eta} - 5\,{x^2}\,{\it \eta} - 4\,{{{\it \eta}}^2} + 
       3\,x\,{{{\it \eta}}^2} \right) +\cr
& & 2\,{\it K_b}{{{\sqrt{{x^2} - 4\,{\it \eta}}}}\over 
   {3{{\left( 1 - x + {\it \eta} \right) }^5}}}\,\left( -5\,{x^2} - 15\,{{{\it \epsilon}}^4}\,{x^2} +
       20\,{{{\it \epsilon}}^6}\,{x^2} + 25\,{x^3} + 21\,{{{\it \epsilon}}^4}\,{x^3} - \right. \cr
& & \left.  10\,{{{\it \epsilon}}^6}\,{x^3} - 50\,{x^4} - 6\,{{{\it \epsilon}}^4}\,{x^4} +
       2\,{{{\it \epsilon}}^6}\,{x^4} + 50\,{x^5} - 25\,{x^6} + 5\,{x^7} +
       14\,{\it \eta} + 6\,{{{\it \epsilon}}^4}\,{\it \eta} \right. \cr
& & \left.       -20\,{{{\it \epsilon}}^6}\,{\it \eta} - 70\,x\,{\it \eta} +
       78\,{{{\it \epsilon}}^4}\,x\,{\it \eta} - 80\,{{{\it \epsilon}}^6}\,x\,{\it \eta} +
       115\,{x^2}\,{\it \eta} - 147\,{{{\it \epsilon}}^4}\,{x^2}\,{\it \eta} + \right. \cr
& & \left.       44\,{{{\it \epsilon}}^6}\,{x^2}\,{\it \eta} - 40\,{x^3}\,{\it \eta} +
       60\,{{{\it \epsilon}}^4}\,{x^3}\,{\it \eta} -
       10\,{{{\it \epsilon}}^6}\,{x^3}\,{\it \eta} - 80\,{x^4}\,{\it \eta} -
       6\,{{{\it \epsilon}}^4}\,{x^4}\,{\it \eta} + 86\,{x^5}\,{\it \eta} \right. \cr
& & \left. -25\,{x^6}\,{\it \eta} + 70\,{{{\it \eta}}^2} -
       126\,{{{\it \epsilon}}^4}\,{{{\it \eta}}^2} +
       152\,{{{\it \epsilon}}^6}\,{{{\it \eta}}^2} - 280\,x\,{{{\it \eta}}^2} +
       300\,{{{\it \epsilon}}^4}\,x\,{{{\it \eta}}^2} -
       80\,{{{\it \epsilon}}^6}\,x\,{{{\it \eta}}^2} \right. \cr
& & \left. + 370\,{x^2}\,{{{\it \eta}}^2} -
       147\,{{{\it \epsilon}}^4}\,{x^2}\,{{{\it \eta}}^2} +
       20\,{{{\it \epsilon}}^6}\,{x^2}\,{{{\it \eta}}^2} -
       130\,{x^3}\,{{{\it \eta}}^2} +
       21\,{{{\it \epsilon}}^4}\,{x^3}\,{{{\it \eta}}^2} -
       80\,{x^4}\,{{{\it \eta}}^2}\right. \cr
& & \left. + 50\,{x^5}\,{{{\it \eta}}^2} +
       140\,{{{\it \eta}}^3} - 126\,{{{\it \epsilon}}^4}\,{{{\it \eta}}^3} -
       20\,{{{\it \epsilon}}^6}\,{{{\it \eta}}^3} - 420\,x\,{{{\it \eta}}^3} +
       78\,{{{\it \epsilon}}^4}\,x\,{{{\it \eta}}^3} + 370\,{x^2}\,{{{\it \eta}}^3} \right. \cr
& & \left. -
       15\,{{{\it \epsilon}}^4}\,{x^2}\,{{{\it \eta}}^3} -
       40\,{x^3}\,{{{\it \eta}}^3} - 50\,{x^4}\,{{{\it \eta}}^3} +
       140\,{{{\it \eta}}^4} + 6\,{{{\it \epsilon}}^4}\,{{{\it \eta}}^4} -
       280\,x\,{{{\it \eta}}^4} + 115\,{x^2}\,{{{\it \eta}}^4} \right. \cr
& & \left. +
       25\,{x^3}\,{{{\it \eta}}^4} + 70\,{{{\it \eta}}^5} -
       70\,x\,{{{\it \eta}}^5} - 5\,{x^2}\,{{{\it \eta}}^5} + 14\,{{{\it \eta}}^6}
        \right),\\
B_W &=& {{\left( {x^2} - 4\,{\it \eta} \right) }
    \over {{{\left( 1 - x + {\it \eta} \right) }^3}}} \,
     {{\left( -1 + {\it \epsilon^2} + x - {\it \eta} \right) }^2}\,
     \left( 1 - {\it \epsilon^2} - 3\,x - {\it \epsilon^2}\,x + 2\,{x^2} + 4\,{\it \eta} + \right. \cr
& & \left.  3\,{\it \epsilon^2}\,{\it \eta} - 5\,x\,{\it \eta} + 3\,{{{\it \eta}}^2} \right) +
   \epsilon_b {{\left( {x^2} - 4\,{\it \eta} \right) }
    \over {{{\left( 1 - x + {\it \eta} \right) }^3}}}\,
     {{\left( -1 + {\it \epsilon^2} + x - {\it \eta} \right) }^2}\,
     \left( 1 - {\it \epsilon^2} - 3\,x - \right. \cr
& & \left.     {\it \epsilon^2}\,x + 2\,{x^2} + 4\,{\it \eta} + 
       3\,{\it \epsilon^2}\,{\it \eta} - 5\,x\,{\it \eta} + 3\,{{{\it \eta}}^2} \right) +
    2\,{\it K_b} {{\left(  -{x^2} + 4\,{\it \eta} \right) }\over 
   {3\,{{\left( 1 - x + {\it \eta} \right) }^5}}}\, 
    \left( 5\,x - \right. \cr
& & \left.    15\,{{{\it \epsilon}}^4}\,x + 10\,{{{\it \epsilon}}^6}\,x -
       25\,{x^2} + 21\,{{{\it \epsilon}}^4}\,{x^2} + 4\,{{{\it \epsilon}}^6}\,{x^2} +
       50\,{x^3} - 6\,{{{\it \epsilon}}^4}\,{x^3} - 2\,{{{\it \epsilon}}^6}\,{x^3} \right. \cr
& & \left. - 50\,{x^4} + 25\,{x^5} - 5\,{x^6} + 36\,{{{\it \epsilon}}^4}\,{\it \eta} -
       36\,{{{\it \epsilon}}^6}\,{\it \eta} + 25\,x\,{\it \eta} -
       51\,{{{\it \epsilon}}^4}\,x\,{\it \eta} - 22\,{{{\it \epsilon}}^6}\,x\,{\it \eta} \right. \cr
& & \left. - 100\,{x^2}\,{\it \eta} + 10\,{{{\it \epsilon}}^6}\,{x^2}\,{\it \eta} +
       150\,{x^3}\,{\it \eta} + 6\,{{{\it \epsilon}}^4}\,{x^3}\,{\it \eta} -
       100\,{x^4}\,{\it \eta} + 25\,{x^5}\,{\it \eta} +
       60\,{{{\it \epsilon}}^6}\,{{{\it \eta}}^2} \right. \cr
& & \left. + 50\,x\,{{{\it \eta}}^2} +
       51\,{{{\it \epsilon}}^4}\,x\,{{{\it \eta}}^2} -
       20\,{{{\it \epsilon}}^6}\,x\,{{{\it \eta}}^2} - 150\,{x^2}\,{{{\it \eta}}^2} -
       21\,{{{\it \epsilon}}^4}\,{x^2}\,{{{\it \eta}}^2} +
       150\,{x^3}\,{{{\it \eta}}^2} \right. \cr
& & \left. - 50\,{x^4}\,{{{\it \eta}}^2} -
       36\,{{{\it \epsilon}}^4}\,{{{\it \eta}}^3} + 50\,x\,{{{\it \eta}}^3} +
       15\,{{{\it \epsilon}}^4}\,x\,{{{\it \eta}}^3} - 100\,{x^2}\,{{{\it \eta}}^3} +
       50\,{x^3}\,{{{\it \eta}}^3} + 25\,x\,{{{\it \eta}}^4}  \right. \cr
& & \left.  -25\,{x^2}\,{{{\it \eta}}^4} + 5\,x\,{{{\it \eta}}^5} \right) .\\
A_H &=&{{{\it \eta}\,{tan^2}\beta \,{\sqrt{{x^2} - 4\,{\it \eta}}}}
    \over {8\,{{{\it \xi}}^4}\,{{\left( 1 - x + {\it \eta} \right) }^3}}}\,
     {{\left( 1 - {{{\it \epsilon}}^2} - x + {\it \eta} \right) }^2}\,
     \left[ L\,\left( 6\,{\it \epsilon}\,x - 6\,{\it \epsilon}\,{x^2} - 
     12\,{\it \epsilon}\,{\it \eta} + \right. \right. \cr
& & \left. \left. 18\,{\it \epsilon}\,x\,{\it \eta} - 
12\,{\it \epsilon}\,{{{\it \eta}}^2} \right)  + 
  F\,\left( 3\,x + 3\,{{{\it \epsilon}}^2}\,x - 5\,{x^2} - {{{\it \epsilon}}^2}\,{x^2} + 
     2\,{x^3} - 4\,{\it \eta} - \right. \right. \cr
& & \left. \left.  8\,{{{\it \epsilon}}^2}\,{\it \eta} + 10\,x\,{\it \eta} + 
     3\,{{{\it \epsilon}}^2}\,x\,{\it \eta} - 5\,{x^2}\,{\it \eta} - 
     4\,{{{\it \eta}}^2} + 3\,x\,{{{\it \eta}}^2} \right) \right] +\cr
& & {\it K_b}\,{{\it \eta}}{{{tan^2 \beta}\,{\sqrt{{x^2} - 4\,{\it \eta}}}}
    \over {12\,{{{\it \xi}}^4}\,{{\left( 1 - x + {\it \eta} \right) }^5}}}\,
     \left[L\,\left( -18\,{\it \epsilon}\,x + 36\,{{{\it \epsilon}}^3}\,x - 18\,{{{\it \epsilon}}^5}\,x + 
     54\,{\it \epsilon}\,{x^2} \right. \right. \cr
& & \left. \left.  - 120\,{{{\it \epsilon}}^3}\,{x^2} + 
     66\,{{{\it \epsilon}}^5}\,{x^2} - 36\,{\it \epsilon}\,{x^3} + 
     132\,{{{\it \epsilon}}^3}\,{x^3} - 60\,{{{\it \epsilon}}^5}\,{x^3} - 
     36\,{\it \epsilon}\,{x^4} \right. \right. \cr 
& & \left. \left. - 48\,{{{\it \epsilon}}^3}\,{x^4} + 
     12\,{{{\it \epsilon}}^5}\,{x^4} + 54\,{\it \epsilon}\,{x^5} - 18\,{\it \epsilon}\,{x^6} + 
     72\,{\it \epsilon}\,{\it \eta} - 96\,{{{\it \epsilon}}^3}\,{\it \eta} + 
     24\,{{{\it \epsilon}}^5}\,{\it \eta} \right. \right. \cr
& &  \left. \left. - 342\,{\it \epsilon}\,x\,{\it \eta} + 
     480\,{{{\it \epsilon}}^3}\,x\,{\it \eta} - 210\,{{{\it \epsilon}}^5}\,x\,{\it \eta} + 
     504\,{\it \epsilon}\,{x^2}\,{\it \eta} - 720\,{{{\it \epsilon}}^3}\,{x^2}\,{\it \eta} + \right. \right. \cr
& & \left. \left.     264\,{{{\it \epsilon}}^5}\,{x^2}\,{\it \eta} - 180\,{\it \epsilon}\,{x^3}\,{\it \eta} + 
     360\,{{{\it \epsilon}}^3}\,{x^3}\,{\it \eta} - 
     60\,{{{\it \epsilon}}^5}\,{x^3}\,{\it \eta} - 144\,{\it \epsilon}\,{x^4}\,{\it \eta} - \right. \right. \cr
& & \left. \left.     24\,{{{\it \epsilon}}^3}\,{x^4}\,{\it \eta} + 90\,{\it \epsilon}\,{x^5}\,{\it \eta} + 
     288\,{\it \epsilon}\,{{{\it \eta}}^2} - 384\,{{{\it \epsilon}}^3}\,{{{\it \eta}}^2} + 
     192\,{{{\it \epsilon}}^5}\,{{{\it \eta}}^2} - 900\,{\it \epsilon}\,x\,{{{\it \eta}}^2}  \right. \right. \cr 
& & \left. \left. 1080\,{{{\it \epsilon}}^3}\,x\,{{{\it \eta}}^2} 
     - 390\,{{{\it \epsilon}}^5}\,x\,{{{\it \eta}}^2} + 
     756\,{\it \epsilon}\,{x^2}\,{{{\it \eta}}^2} - 
     792\,{{{\it \epsilon}}^3}\,{x^2}\,{{{\it \eta}}^2} + 
     102\,{{{\it \epsilon}}^5}\,{x^2}\,{{{\it \eta}}^2} \right. \right. \cr
& & \left. \left. + 
     36\,{\it \epsilon}\,{x^3}\,{{{\it \eta}}^2} + 
     84\,{{{\it \epsilon}}^3}\,{x^3}\,{{{\it \eta}}^2} - 
     180\,{\it \epsilon}\,{x^4}\,{{{\it \eta}}^2} + 432\,{\it \epsilon}\,{{{\it \eta}}^3} - 
     480\,{{{\it \epsilon}}^3}\,{{{\it \eta}}^3} + 
     168\,{{{\it \epsilon}}^5}\,{{{\it \eta}}^3} \right. \right. \cr
& & \left. \left. - 828\,{\it \epsilon}\,x\,{{{\it \eta}}^3} + 
     672\,{{{\it \epsilon}}^3}\,x\,{{{\it \eta}}^3} - 
     54\,{{{\it \epsilon}}^5}\,x\,{{{\it \eta}}^3} + 
     216\,{\it \epsilon}\,{x^2}\,{{{\it \eta}}^3} - 
     96\,{{{\it \epsilon}}^3}\,{x^2}\,{{{\it \eta}}^3} + \right. \right. \cr
& & \left. \left.     180\,{\it \epsilon}\,{x^3}\,{{{\it \eta}}^3} + 288\,{\it \epsilon}\,{{{\it \eta}}^4} - 
     192\,{{{\it \epsilon}}^3}\,{{{\it \eta}}^4} - 234\,{\it \epsilon}\,x\,{{{\it \eta}}^4} + 
     36\,{{{\it \epsilon}}^3}\,x\,{{{\it \eta}}^4} - 
     90\,{\it \epsilon}\,{x^2}\,{{{\it \eta}}^4} \right. \right. \cr
& & \left. \left.  + 72\,{\it \epsilon}\,{{{\it \eta}}^5} + 
     18\,{\it \epsilon}\,x\,{{{\it \eta}}^5} \right)  + 
  F\,\left( -5\,{x^2} - 15\,{{{\it \epsilon}}^4}\,{x^2} + 
     20\,{{{\it \epsilon}}^6}\,{x^2} + 25\,{x^3} + 21\,{{{\it \epsilon}}^4}\,{x^3} \right. \right. \cr
& & \left. \left. - 
     10\,{{{\it \epsilon}}^6}\,{x^3} - 50\,{x^4} - 6\,{{{\it \epsilon}}^4}\,{x^4} + 
     2\,{{{\it \epsilon}}^6}\,{x^4} + 50\,{x^5} - 25\,{x^6} + 5\,{x^7} + 
     14\,{\it \eta} + \right. \right. \cr
& & \left. \left. 6\,{{{\it \epsilon}}^4}\,{\it \eta} - 
     20\,{{{\it \epsilon}}^6}\,{\it \eta} - 70\,x\,{\it \eta} + 
     78\,{{{\it \epsilon}}^4}\,x\,{\it \eta} - 80\,{{{\it \epsilon}}^6}\,x\,{\it \eta} + 
     115\,{x^2}\,{\it \eta} - 147\,{{{\it \epsilon}}^4}\,{x^2}\,{\it \eta} \right. \right. \cr
& & \left. \left.    + 
     44\,{{{\it \epsilon}}^6}\,{x^2}\,{\it \eta} - 40\,{x^3}\,{\it \eta}  + 
     60\,{{{\it \epsilon}}^4}\,{x^3}\,{\it \eta} - 
     10\,{{{\it \epsilon}}^6}\,{x^3}\,{\it \eta} - 80\,{x^4}\,{\it \eta} - 
     6\,{{{\it \epsilon}}^4}\,{x^4}\,{\it \eta} + \right. \right. \cr
& & \left. \left. 86\,{x^5}\,{\it \eta} - 
     25\,{x^6}\,{\it \eta} + 70\,{{{\it \eta}}^2} - 
     126\,{{{\it \epsilon}}^4}\,{{{\it \eta}}^2} + 
     152\,{{{\it \epsilon}}^6}\,{{{\it \eta}}^2} - 280\,x\,{{{\it \eta}}^2} + 
     300\,{{{\it \epsilon}}^4}\,x\,{{{\it \eta}}^2} \right. \right. \cr
& & \left. \left.     - 
     80\,{{{\it \epsilon}}^6}\,x\,{{{\it \eta}}^2} + 370\,{x^2}\,{{{\it \eta}}^2} - 
     147\,{{{\it \epsilon}}^4}\,{x^2}\,{{{\it \eta}}^2} + 
     20\,{{{\it \epsilon}}^6}\,{x^2}\,{{{\it \eta}}^2} - 130\,{x^3}\,{{{\it \eta}}^2} + 
     21\,{{{\it \epsilon}}^4}\,{x^3}\,{{{\it \eta}}^2} \right. \right. \cr
& & \left. \left.     - 80\,{x^4}\,{{{\it \eta}}^2} + 
     50\,{x^5}\,{{{\it \eta}}^2} + 140\,{{{\it \eta}}^3} - 
     126\,{{{\it \epsilon}}^4}\,{{{\it \eta}}^3} - 
     20\,{{{\it \epsilon}}^6}\,{{{\it \eta}}^3} - 420\,x\,{{{\it \eta}}^3} + 
     78\,{{{\it \epsilon}}^4}\,x\,{{{\it \eta}}^3} \right. \right. \cr
& & \left. \left.     + 370\,{x^2}\,{{{\it \eta}}^3} - 
     15\,{{{\it \epsilon}}^4}\,{x^2}\,{{{\it \eta}}^3} - 40\,{x^3}\,{{{\it \eta}}^3} - 
     50\,{x^4}\,{{{\it \eta}}^3} + 140\,{{{\it \eta}}^4} + 
     6\,{{{\it \epsilon}}^4}\,{{{\it \eta}}^4} - 280\,x\,{{{\it \eta}}^4} \right. \right. \cr 
& & \left. \left.    + 115\,{x^2}\,{{{\it \eta}}^4} + 25\,{x^3}\,{{{\it \eta}}^4} + 
     70\,{{{\it \eta}}^5} - 70\,x\,{{{\it \eta}}^5} - 5\,{x^2}\,{{{\it \eta}}^5} + 
     14\,{{{\it \eta}}^6} \right) \right],\\
B_H &=&-\frac{J\,r^2\,{\it \eta}}{8\,\xi^2}\,B_W,\\
A_I &=&-{{3\,{\it \eta}\,{\sqrt{{x^2} - 4\,{\it \eta}}}\,
     {{\left( -1 + {{{\it \epsilon}}^2} + x - {\it \eta} \right) }^2}}\over 
   {{{{\it \xi}}^2}\,{{\left( 1 - x + {\it \eta} \right) }^2}}}\,
     \left( 2\,{{{\it \epsilon}}^2} + 2\,{{{\it tan}}^2 \beta} - 
     {{{\it \epsilon}}^2}\,x - 
       2\,x\,{{{\it tan}}^2 \beta} +  \right. \cr
& &  \left.  2\,{{{\it tan}}^2 \beta}\,{\it \eta} \right) - 
   {\it K_b}{{{\it \eta}\,{\sqrt{{x^2} - 
   4\,{\it \eta}}}}\over 
   {{{{\it \xi}}^2}\,{{\left( 1 - x + {\it \eta} \right) }^4}}}\left[6\,{{{\it \epsilon}}^2} - 12\,{{{\it \epsilon}}^4} + 6\,{{{\it \epsilon}}^6} - 
  24\,{{{\it \epsilon}}^2}\,x + 24\,{{{\it \epsilon}}^4}\,x + \right. \cr
& & \left. 36\,{{{\it \epsilon}}^2}\,{x^2} - 
  14\,{{{\it \epsilon}}^4}\,{x^2} - 24\,{{{\it \epsilon}}^2}\,{x^3} + 
  6\,{{{\it \epsilon}}^2}\,{x^4} + 2\,{{{\it \epsilon}}^4}\,{x^4} + 
  24\,{{{\it \epsilon}}^2}\,{\it \eta} - 4\,{{{\it \epsilon}}^4}\,{\it \eta} - 
  36\,{{{\it \epsilon}}^6}\,{\it \eta} \right. \cr
& & \left.  - 72\,{{{\it \epsilon}}^2}\,x\,{\it \eta} + 
  24\,{{{\it \epsilon}}^6}\,x\,{\it \eta} + 72\,{{{\it \epsilon}}^2}\,{x^2}\,{\it \eta} + 
  8\,{{{\it \epsilon}}^4}\,{x^2}\,{\it \eta} - 6\,{{{\it \epsilon}}^6}\,{x^2}\,{\it \eta} - 
  24\,{{{\it \epsilon}}^2}\,{x^3}\,{\it \eta} - 8\,{{{\it \epsilon}}^4}\,{x^3}\,{\it \eta}  \right.\cr
& & \left. +36\,{{{\it \epsilon}}^2}\,{{{\it \eta}}^2} - 4\,{{{\it \epsilon}}^4}\,{{{\it \eta}}^2} + 
  6\,{{{\it \epsilon}}^6}\,{{{\it \eta}}^2} - 72\,{{{\it \epsilon}}^2}\,x\,{{{\it \eta}}^2} + 
  8\,{{{\it \epsilon}}^4}\,x\,{{{\it \eta}}^2} + 
  36\,{{{\it \epsilon}}^2}\,{x^2}\,{{{\it \eta}}^2} + 
  6\,{{{\it \epsilon}}^4}\,{x^2}\,{{{\it \eta}}^2} \right. \cr
& & \left. + 
  24\,{{{\it \epsilon}}^2}\,{{{\it \eta}}^3} - 12\,{{{\it \epsilon}}^4}\,{{{\it \eta}}^3} - 
  24\,{{{\it \epsilon}}^2}\,x\,{{{\it \eta}}^3} + 6\,{{{\it \epsilon}}^2}\,{{{\it \eta}}^4} + 
  {{{\it tan}}^2 \beta}\,\left( -6 + 12\,{{{\it \epsilon}}^2} - 6\,{{{\it \epsilon}}^4} + 
     24\,x \right. \right. \cr
& & \left. \left. - 48\,{{{\it \epsilon}}^2}\,x + 24\,{{{\it \epsilon}}^4}\,x - 36\,{x^2} + 
     72\,{{{\it \epsilon}}^2}\,{x^2} - 26\,{{{\it \epsilon}}^4}\,{x^2} + 24\,{x^3} - 
     48\,{{{\it \epsilon}}^2}\,{x^3} + 8\,{{{\it \epsilon}}^4}\,{x^3} \right. \right. \cr
& & \left. \left. - 6\,{x^4} + 
     12\,{{{\it \epsilon}}^2}\,{x^4} - 30\,{\it \eta} + 
     48\,{{{\it \epsilon}}^2}\,{\it \eta} - 34\,{{{\it \epsilon}}^4}\,{\it \eta} + 
     96\,x\,{\it \eta} - 144\,{{{\it \epsilon}}^2}\,x\,{\it \eta} + 
     64\,{{{\it \epsilon}}^4}\,x\,{\it \eta} \right. \right. \cr
& & \left. \left.     - 108\,{x^2}\,{\it \eta} + 
     144\,{{{\it \epsilon}}^2}\,{x^2}\,{\it \eta} - 
     26\,{{{\it \epsilon}}^4}\,{x^2}\,{\it \eta} + 48\,{x^3}\,{\it \eta} - 
     48\,{{{\it \epsilon}}^2}\,{x^3}\,{\it \eta} - 6\,{x^4}\,{\it \eta} - 
     60\,{{{\it \eta}}^2} \right. \right. \cr
& & \left. \left. + 72\,{{{\it \epsilon}}^2}\,{{{\it \eta}}^2} - 
     34\,{{{\it \epsilon}}^4}\,{{{\it \eta}}^2} + 144\,x\,{{{\it \eta}}^2} - 
     144\,{{{\it \epsilon}}^2}\,x\,{{{\it \eta}}^2} + 
     24\,{{{\it \epsilon}}^4}\,x\,{{{\it \eta}}^2} - 108\,{x^2}\,{{{\it \eta}}^2} + \right.\right. \cr
& & \left. \left.     72\,{{{\it \epsilon}}^2}\,{x^2}\,{{{\it \eta}}^2} + 24\,{x^3}\,{{{\it \eta}}^2} - 
     60\,{{{\it \eta}}^3} + 48\,{{{\it \epsilon}}^2}\,{{{\it \eta}}^3} - 
     6\,{{{\it \epsilon}}^4}\,{{{\it \eta}}^3} + 96\,x\,{{{\it \eta}}^3} - 
     48\,{{{\it \epsilon}}^2}\,x\,{{{\it \eta}}^3} - \right. \right. \cr
& & \left. \left. 36\,{x^2}\,{{{\it \eta}}^3} - 
     30\,{{{\it \eta}}^4} + 12\,{{{\it \epsilon}}^2}\,{{{\it \eta}}^4} + 
     24\,x\,{{{\it \eta}}^4} - 6\,{{{\it \eta}}^5} \right) \right]  ,\\
B_I &=&{{3\,{\it \eta}\,{{{\it \epsilon}}^2}\,{{\left( -1 + 
{{{\it \epsilon}}^2} + x - {\it \eta} \right) }^
       2}\,\left( {x^2} - 4\,{\it \eta} \right) }\over 
   {{{{\it \xi}}^2}\,{{\left( 1 - x + {\it \eta} \right) }^2}}}+ 
  \epsilon_b \,
   {{3\,{\it \eta}\,{{{\it \epsilon}}^2}\,{{\left( -1 + 
   {{{\it \epsilon}}^2} + x - {\it \eta} \right) }^
       2}\,\left( {x^2} - 4\,{\it \eta} \right) }\over 
   {{{{\it \xi}}^2}\,{{\left( 1 - x + {\it \eta} \right) }^2}}} \cr
& &  -{\it K_b}{{{\it \eta}\,\left( 1 - 
   {{{\it \epsilon}}^2} - x + {\it \eta} \right) }\over 
   {{{{\it \xi}}^2}\,{{\left( 1 - x + {\it \eta} \right) }^4}}}\left[-22\,{{{\it \epsilon}}^2}\,x + 20\,{{{\it \epsilon}}^4}\,x + 2\,{{{\it \epsilon}}^6}\,x + 
  62\,{{{\it \epsilon}}^2}\,{x^2} - 30\,{{{\it \epsilon}}^4}\,{x^2} - \right. \cr
& & \left.  8\,{{{\it \epsilon}}^6}\,{x^2} - 52\,{{{\it \epsilon}}^2}\,{x^3} + 
  4\,{{{\it \epsilon}}^4}\,{x^3} + 4\,{{{\it \epsilon}}^2}\,{x^4} + 
  6\,{{{\it \epsilon}}^4}\,{x^4} + 10\,{{{\it \epsilon}}^2}\,{x^5} - 
  2\,{{{\it \epsilon}}^2}\,{x^6} + 92\,{{{\it \epsilon}}^2}\,{\it \eta} - \right. \cr
& & \left.  104\,{{{\it \epsilon}}^4}\,{\it \eta} + 12\,{{{\it \epsilon}}^6}\,{\it \eta} - 
  360\,{{{\it \epsilon}}^2}\,x\,{\it \eta} + 236\,{{{\it \epsilon}}^4}\,x\,{\it \eta} + 
  20\,{{{\it \epsilon}}^6}\,x\,{\it \eta} + 448\,{{{\it \epsilon}}^2}\,{x^2}\,{\it \eta} - 
  126\,{{{\it \epsilon}}^4}\,{x^2}\,{\it \eta} \right. \cr
& & \left. + 8\,{{{\it \epsilon}}^6}\,{x^2}\,{\it \eta} - 
  176\,{{{\it \epsilon}}^2}\,{x^3}\,{\it \eta} - 4\,{{{\it \epsilon}}^4}\,{x^3}\,{\it \eta} - 
  2\,{{{\it \epsilon}}^6}\,{x^3}\,{\it \eta} - 12\,{{{\it \epsilon}}^2}\,{x^4}\,{\it \eta} + 
  2\,{{{\it \epsilon}}^4}\,{x^4}\,{\it \eta} + 8\,{{{\it \epsilon}}^2}\,{x^5}\,{\it \eta}  \right. \cr
& & \left. +  272\,{{{\it \epsilon}}^2}\,{{{\it \eta}}^2} - 152\,{{{\it \epsilon}}^4}\,{{{\it \eta}}^2} - 
  72\,{{{\it \epsilon}}^6}\,{{{\it \eta}}^2} - 
  660\,{{{\it \epsilon}}^2}\,x\,{{{\it \eta}}^2} + 
  156\,{{{\it \epsilon}}^4}\,x\,{{{\it \eta}}^2} + 
  18\,{{{\it \epsilon}}^6}\,x\,{{{\it \eta}}^2} + \right. \cr
& & \left.  420\,{{{\it \epsilon}}^2}\,{x^2}\,{{{\it \eta}}^2} + 
  6\,{{{\it \epsilon}}^4}\,{x^2}\,{{{\it \eta}}^2} - 
  4\,{{{\it \epsilon}}^6}\,{x^2}\,{{{\it \eta}}^2} - 
  20\,{{{\it \epsilon}}^2}\,{x^3}\,{{{\it \eta}}^2} - 
  8\,{{{\it \epsilon}}^4}\,{x^3}\,{{{\it \eta}}^2} - 
  12\,{{{\it \epsilon}}^2}\,{x^4}\,{{{\it \eta}}^2} + \right. \cr 
& & \left.  264\,{{{\it \epsilon}}^2}\,{{{\it \eta}}^3} - 56\,{{{\it \epsilon}}^4}\,{{{\it \eta}}^3} + 
  12\,{{{\it \epsilon}}^6}\,{{{\it \eta}}^3} - 
  328\,{{{\it \epsilon}}^2}\,x\,{{{\it \eta}}^3} + 
  4\,{{{\it \epsilon}}^4}\,x\,{{{\it \eta}}^3} + 
  32\,{{{\it \epsilon}}^2}\,{x^2}\,{{{\it \eta}}^3} + 
  6\,{{{\it \epsilon}}^4}\,{x^2}\,{{{\it \eta}}^3} \right. \cr
& & \left. + 
  8\,{{{\it \epsilon}}^2}\,{x^3}\,{{{\it \eta}}^3} + 
  80\,{{{\it \epsilon}}^2}\,{{{\it \eta}}^4} - 8\,{{{\it \epsilon}}^4}\,{{{\it \eta}}^4} - 
  6\,{{{\it \epsilon}}^2}\,x\,{{{\it \eta}}^4} - 
  2\,{{{\it \epsilon}}^2}\,{x^2}\,{{{\it \eta}}^4} - 
  4\,{{{\it \epsilon}}^2}\,{{{\it \eta}}^5} + 
  {{{\it tan}}^2 \beta}\,\left( 2\,x - \right. \right. \cr
& & \left.\left. 4\,{{{\it \epsilon}}^2}\,x + 
     2\,{{{\it \epsilon}}^4}\,x - 6\,{x^2} + 10\,{{{\it \epsilon}}^2}\,{x^2} - 
     4\,{{{\it \epsilon}}^4}\,{x^2} + 4\,{x^3} - 6\,{{{\it \epsilon}}^2}\,{x^3} + 
     2\,{{{\it \epsilon}}^4}\,{x^3} + 4\,{x^4} - \right. \right. \cr
& & \left. \left. 2\,{{{\it \epsilon}}^2}\,{x^4} - 
     6\,{x^5} + 2\,{{{\it \epsilon}}^2}\,{x^5} + 2\,{x^6} - 16\,{\it \eta} + 
     16\,{{{\it \epsilon}}^2}\,{\it \eta} + 74\,x\,{\it \eta} - 
     60\,{{{\it \epsilon}}^2}\,x\,{\it \eta} + 2\,{{{\it \epsilon}}^4}\,x\,{\it \eta} \right. \right. \cr
& & \left. \left. - 
     120\,{x^2}\,{\it \eta} + 66\,{{{\it \epsilon}}^2}\,{x^2}\,{\it \eta} + 
     76\,{x^3}\,{\it \eta} - 16\,{{{\it \epsilon}}^2}\,{x^3}\,{\it \eta} - 
     2\,{{{\it \epsilon}}^4}\,{x^3}\,{\it \eta} - 8\,{x^4}\,{\it \eta} - 
     6\,{{{\it \epsilon}}^2}\,{x^4}\,{\it \eta} - \right. \right. \cr
& & \left. \left. 6\,{x^5}\,{\it \eta} - 
     64\,{{{\it \eta}}^2} + 48\,{{{\it \epsilon}}^2}\,{{{\it \eta}}^2} + 
     212\,x\,{{{\it \eta}}^2} - 108\,{{{\it \epsilon}}^2}\,x\,{{{\it \eta}}^2} - 
     2\,{{{\it \epsilon}}^4}\,x\,{{{\it \eta}}^2} - 228\,{x^2}\,{{{\it \eta}}^2} + \right. \right. \cr
& & \left. \left.      54\,{{{\it \epsilon}}^2}\,{x^2}\,{{{\it \eta}}^2} + 
     4\,{{{\it \epsilon}}^4}\,{x^2}\,{{{\it \eta}}^2} + 76\,{x^3}\,{{{\it \eta}}^2} + 
     6\,{{{\it \epsilon}}^2}\,{x^3}\,{{{\it \eta}}^2} + 4\,{x^4}\,{{{\it \eta}}^2} - 
     96\,{{{\it \eta}}^3} + 48\,{{{\it \epsilon}}^2}\,{{{\it \eta}}^3} +\right. \right. \cr 
& & \left. \left.     212\,x\,{{{\it \eta}}^3} - 52\,{{{\it \epsilon}}^2}\,x\,{{{\it \eta}}^3} - 
     2\,{{{\it \epsilon}}^4}\,x\,{{{\it \eta}}^3} - 120\,{x^2}\,{{{\it \eta}}^3} - 
     2\,{{{\it \epsilon}}^2}\,{x^2}\,{{{\it \eta}}^3} + 4\,{x^3}\,{{{\it \eta}}^3} - 
     64\,{{{\it \eta}}^4} + \right.\right. \cr
& & \left. \left. 16\,{{{\it \epsilon}}^2}\,{{{\it \eta}}^4} + 
     74\,x\,{{{\it \eta}}^4} - 6\,{x^2}\,{{{\it \eta}}^4} - 16\,{{{\it \eta}}^5} + 
     2\,x\,{{{\it \eta}}^5} \right) \right], 
\end{eqnarray}
where
$$F=2tan^2\beta+2\epsilon^2 cot^2\beta,J=2tan^2\beta-2\epsilon^2 
cot^2\beta,L=4\epsilon.$$

 When $\tan\beta\gg 1$, eqs. (3), (4) and (5) reduce to
\begin{eqnarray}
A_H &=& \frac{r^4\,\eta}{4}\,A_W+O(r^4\,\tan\beta^{-2}),\\
B_H &=& -\frac{r^4\,\eta}{4}\,B_W+O(r^4\,tan\beta^{-4}),\\
A_I &=&
-\frac{6\,\eta\,r^2\,\sqrt{x^2-4\,\eta}}{(1-x+\eta)}\,(1-x+\eta-\epsilon^2)^2
+O(r^2\,\tan\beta^{-2}).
\end{eqnarray}

\begin{figure}
\caption{ Total width (normalized to the electron channel) in terms of $tan\beta$ and $m_H$, using the 
(a)first $(m_b=5.044Gev)$,(b)second $(m_b=5.1Gev)$ set of value. The curves terminating
at $tan\beta=100,150,200$ correspond to $m_H=200Gev, 300Gev, 400Gev$ respectively.
 }
\end{figure}
\begin{figure}
\caption{$\tau$ spectrum for different $\alpha_s$ and 
$tan{\beta}$, $m_H=200Gev$. The first set of parameter value$(m_b=5.044Gev)$ is
used. }
\label{f2}
\end{figure}
\begin{figure}
\caption{$\tau$ spectrum for different $\alpha_s$ and 
$tan{\beta}$, $m_H=200Gev$. The second set of parameter value$(m_{b}=5.1Gev)$ is
used.
}
\end{figure}
\begin{figure}
\caption{$\tau$ spectrum for different 
$tan{\beta}$, $m_H$=200Gev. The first set of parameter value$(m_{b}=5.044Gev)$ is
used. $O(\alpha_{s})$ corrections are not considered here.
 }
\end{figure}
\begin{figure}
\caption{$\tau$ spectrum for different  
$tan{\beta}$, $m_H$=200Gev. The second set of parameter value$(m_{b}=5.1Gev)$ is
used. $O(\alpha_{s})$ corrections are not considered here.
 }
\end{figure}

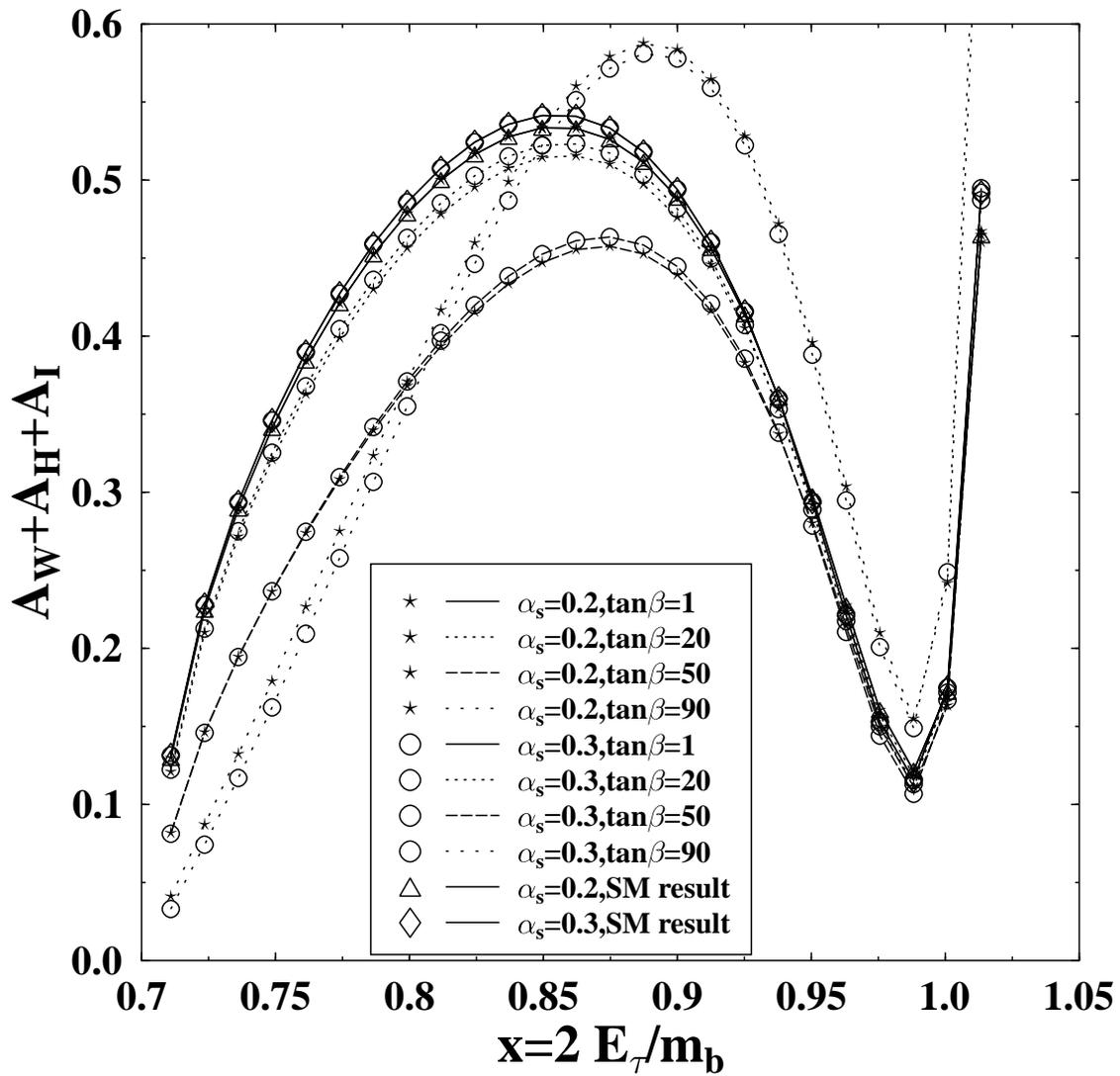
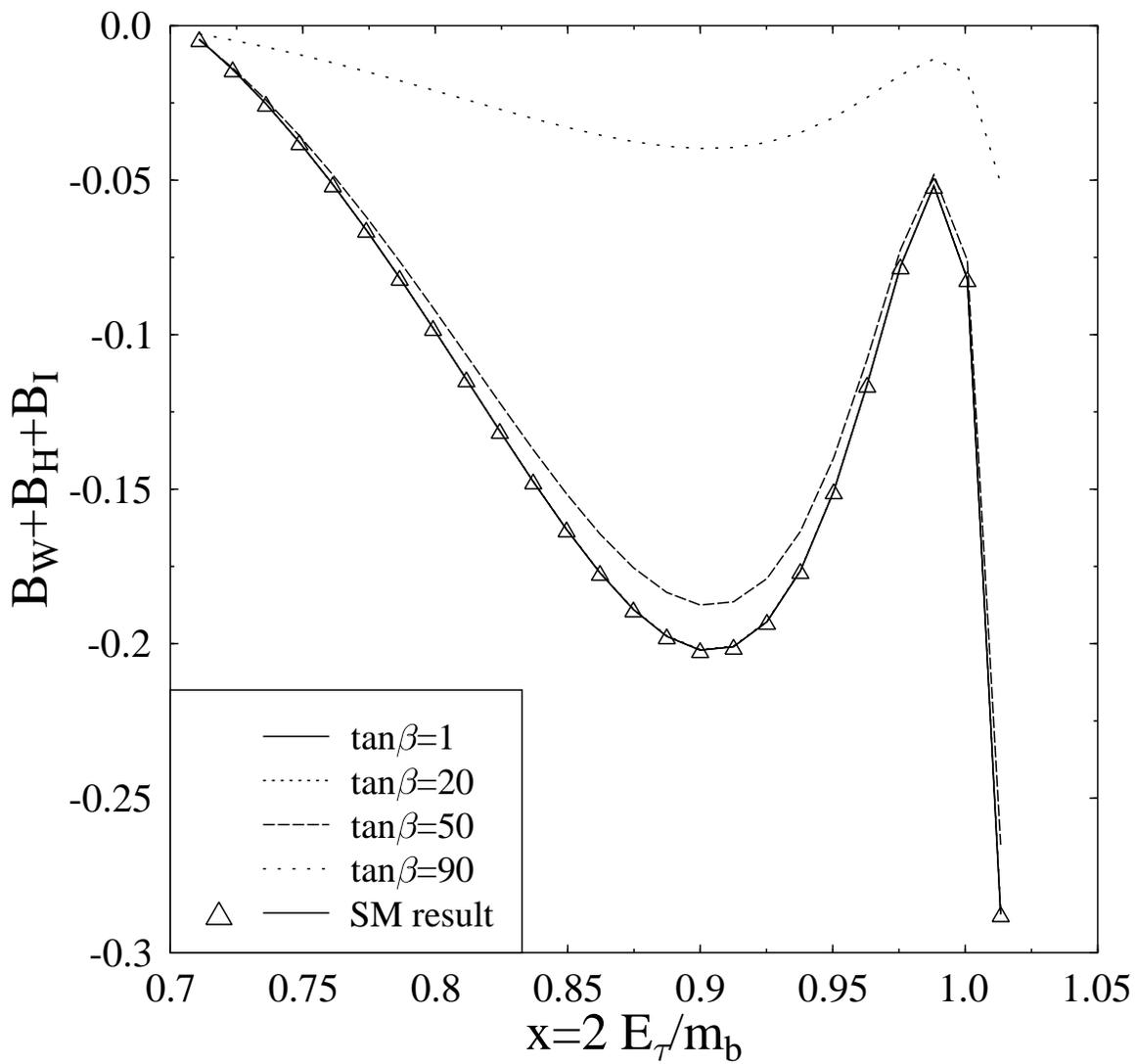
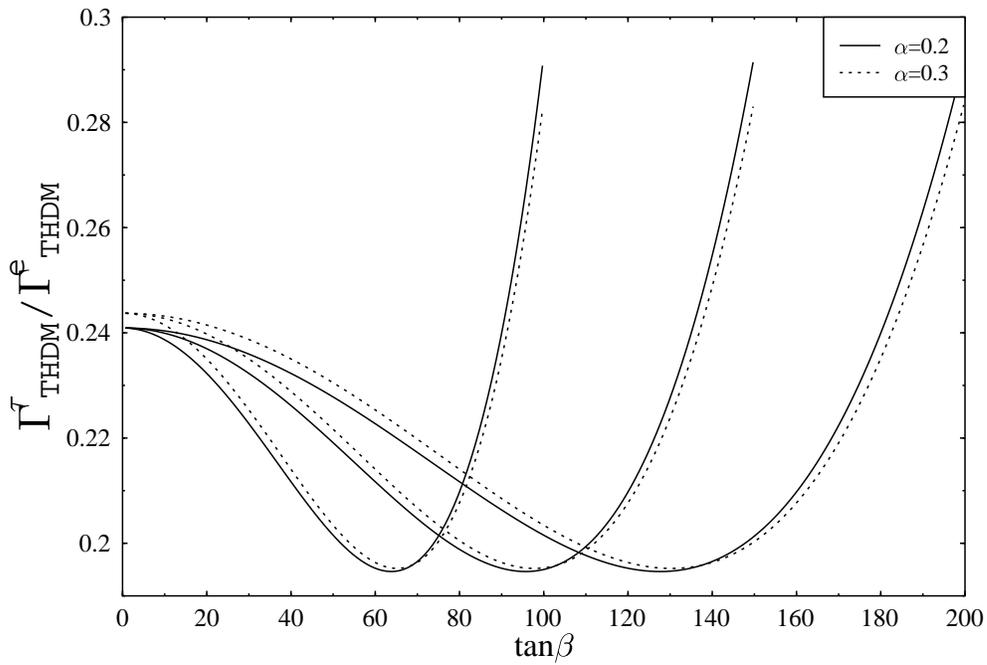
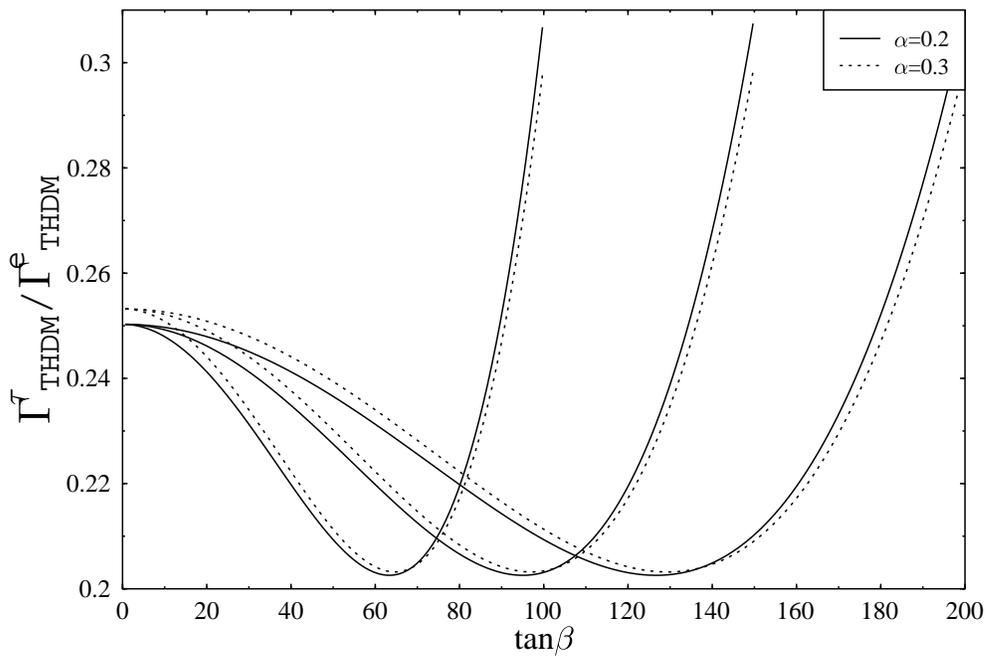
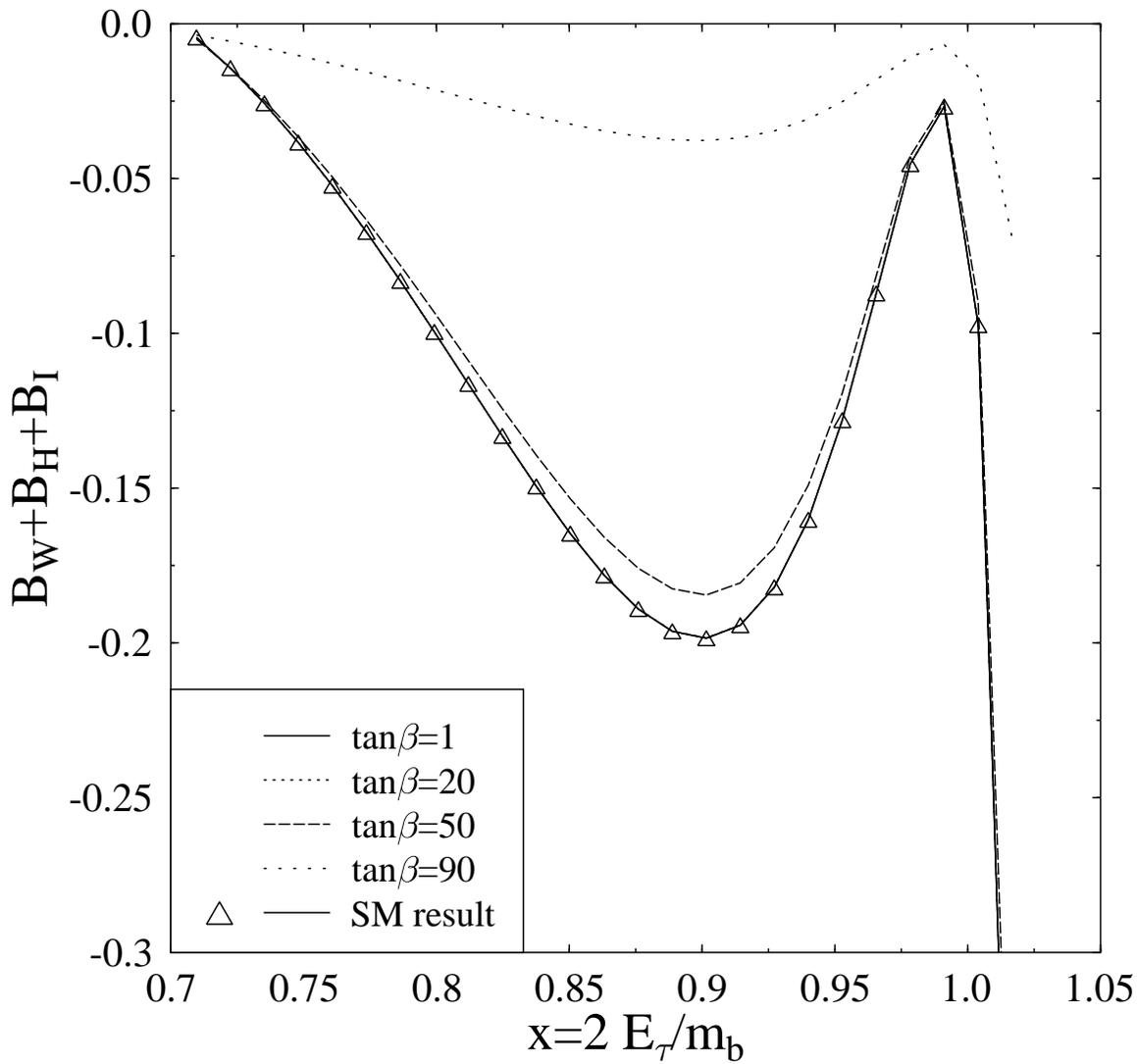
\end{document}